\documentclass[prb,aps,twocolumn,amsmath,amssymb,floatfix,superscriptaddress]{revtex4}

\usepackage[dvips]{graphics}
\usepackage{color}
\usepackage{bbold}
\usepackage{physics}
\usepackage{soul}
\definecolor{dred}{rgb}{0.5,0.2,0}
\usepackage[colorlinks=true, citecolor=blue, urlcolor=blue]{hyperref}
\usepackage{tabularx}
\usepackage{graphicx} 
\date{\today}

\begin{document}

\title{Energy-Selective Complete Spin Polarization in an Extended Su-Schrieffer-Heeger Ferromagnetic Chain}








\author{Souvik Roy}
\email{souvikroy138@gmail.com}
\affiliation{School of Physical Sciences, National Institute of Science Education and Research, Jatni 752050, India}
\affiliation{Homi Bhabha National Institute, Training School Complex, Anushaktinagar, Mumbai 400094, India}

\author{Ranjini Bhattacharya}
\email{ranjinibhattacharya@gmail.com}
\affiliation{Institute of Physics, Sachivalaya Marg, Bhubaneswar-751005, India}
\affiliation{Homi Bhabha National Institute, Training School Complex, Anushaktinagar, Mumbai 400094, India}

\begin{abstract}

We study spin-dependent transport in an extended Su–Schrieffer–Heeger chain with cosine-modulated nearest- and next-nearest-neighbor hopping using the nonequilibrium Green’s function formalism. Suitable tuning of the hopping parameters yields a complete separation of spin channels and perfect spin polarization over broad energy windows. The inclusion of next-nearest-neighbor hopping enhances both tunability and robustness, while systematic phase-diagram analyses reveal quantized polarization ($P_{\sigma}=\pm 1$) across extended regions of parameter space rather than at isolated fine-tuned points. These characteristics persist for larger system sizes, establishing the extended SSH model as a versatile platform for controllable spin-polarized transport.
\end{abstract}

\maketitle


 \paragraph*{Introduction.-} 
Spin-dependent transport in low-dimensional systems lies at the heart of spintronics and quantum-information technologies~\cite{c1,c2,c3,c4,c5,c6,c7,c8,c9,c10,c11}, enabling key functionalities such as spin filtering, spin transistors, and solid-state qubits~\cite{c12,c13}. However, conventional approaches relying on ferromagnetic contacts or external magnetic fields encounter fundamental limitations at the nanoscale, thereby motivating the pursuit of alternative, field-free strategies for achieving robust and controllable spin polarization.

A variety of strategies have been explored to address the challenge of generating and manipulating spin-polarized currents in low-dimensional systems. Prominent among them are spin–orbit (SO) coupled platforms~\cite{c14,c15,c16,c17,c18,c19}, including Rashba~\cite{c20} and Dresselhaus~\cite{c21} interactions, which enable spin polarization from initially unpolarized carriers. Rashba SO coupling is particularly appealing owing to its tunability via external gate voltages, offering controllable spin-selective transport, whereas Dresselhaus coupling is fixed by the bulk crystal symmetry. Nonetheless, the typically weak strength of SO interactions limits the achievable separation between spin-resolved transport channels, posing a key obstacle to efficient spin filtering.

Magnetic materials provide an alternative route to spin-dependent transport through spin–spin exchange interactions~\cite{c22,c23,c24,c25}, which are typically much stronger and can reach energy scales of several electron volts. As a result, magnetic systems with finite magnetization have long served as key building blocks in spintronic devices~\cite{c26,c27,c28,c29,c30}. Spin filtering in such platforms is commonly realized by controlling the relative alignment of magnetic moments via external magnetic fields. However, this approach becomes increasingly impractical at the nanoscale, where the generation and confinement of even moderate magnetic fields present significant experimental challenges.

These limitations motivate the pursuit of alternative, nonmagnetic routes for achieving efficient and controllable spin polarization. In this work, we propose and systematically investigate a robust mechanism for spin-polarized transport in an extended Su–Schrieffer–Heeger (SSH) chain symmetrically coupled to metallic leads~\cite{ssh1,ssh2,ssh3,ssh4,ssh5,ssh6,ssh7}. Quasiperiodicity is incorporated through Fibonacci-modulated onsite energies~\cite{fb1,fb2}, while both nearest-neighbor (NN) and next-nearest-neighbor (NNN) hopping processes with cosine-modulated amplitudes are included. Employing the nonequilibrium Green’s function formalism~\cite{negf1,negf2} at zero temperature, we demonstrate that suitable tuning of the hopping parameters induces a complete separation of up- and down-spin transmission channels, yielding perfect spin polarization over broad energy windows. Comprehensive phase-diagram analyses reveal that quantized spin polarization, 
$P_{\sigma}=\pm 1$, emerges robustly across extended regions of parameter space, rather than at isolated fine-tuned points. Moreover, the distinct roles of intracell and intercell hopping processes enable enhanced flexibility and control over spin transport. The persistence of perfect spin polarization with increasing system size further confirms the intrinsic robustness of the proposed mechanism. Taken together, our results establish the extended SSH model with quasiperiodicity and modulated long-range hopping as a versatile and scalable platform for controllable spin-polarized transport in low-dimensional systems.

\paragraph*{Model.-}
The central ferromagnetic (FM) chain consists of atomic sites hosting localized magnetic moments, whose orientations are generally described by the polar and azimuthal angles $\theta_i$ and $\psi_i$. In this work, we consider a simplified yet physically relevant configuration in which all local moments are rigidly aligned along the $+Z$ direction, thereby stabilizing a uniform ferromagnetic phase with complete spin polarization and a net magnetization oriented along $+Z$. This idealized, fluctuation-free background provides a transparent platform for investigating how an intrinsically spin-polarized environment governs spin-dependent transport~\cite{rb1,rb2,rb3}.

\begin{figure}[ht]
{\centering \resizebox*{7.8cm}{3.2cm}{\includegraphics{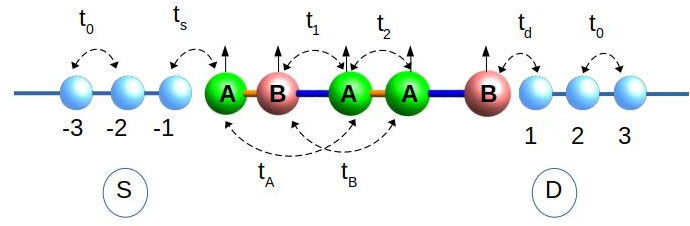}}\par}
\caption{(Color online). The schematic illustrates a long-range Su-Schrieffer-Heeger (SSH) chain symmetrically coupled to source and drain electrodes, with onsite energies modulated according to a Fibonacci sequence. The interplay between quasiperiodicity and long-range hopping provides a versatile platform for spin-polarized transport.}
\label{schematic}
\end{figure}

The total Hamiltonian of the system is written as
\begin{equation}
H= H_L + H_S + H_D + H_{tn},
\label{eq2}
\end{equation}
where $H_L$ describes the central SSH chain, $H_S$ and $H_D$ correspond to the source and drain electrodes, respectively, and $H_{tn}$ accounts for the tunneling coupling between the electrodes and the chain.

The Hamiltonian of the central region reads
\textcolor{black}{
\begin{align}
H_L &= \sum_n \hat{c}_n^\dagger 
       \big(\epsilon_n \otimes \mathbb{I}_1 - \mathbf{h}_n \cdot \boldsymbol{\sigma} \big) \hat{c}_n \nonumber\\
    &\quad + \sum_{n = odd } \Big(
       \hat{c}_n^\dagger \, \hat{t}_1 \, \hat{c}_{n+1} 
       + \mathrm{H.c.} \Big) \nonumber\\
    &\quad + \sum_{n=even} \Big(
       \hat{c}_{n}^\dagger \, \hat{t}_2 \, \hat{c}_{n+1} 
       + \mathrm{H.c.} \Big) \nonumber\\
    &\quad + \sum_{n = odd} \Big(
       \hat{c}_n^\dagger \, \hat{t}_A \, \hat{c}_{n+2} 
       + \mathrm{H.c.} \Big) \nonumber\\
    &\quad + \sum_{n = even} \Big(
       \hat{c}_{n}^\dagger \, \hat{t}_B \, \hat{c}_{n+2} 
       + \mathrm{H.c.} \Big).
\end{align}
}

Here $\hat{c}_n=(c_{\uparrow,n},\,c_{\downarrow,n})^{T}$ is the spinor operator at site $n$. The onsite potential $\epsilon_n$ is a $2\times2$ matrix in spin space, while $\mathbb{I}_1$ denotes the identity matrix. The local spin field $\mathbf{h}_n\cdot\boldsymbol{\sigma}$ describes the coupling to uniformly aligned magnetic moments taken along the $+Z$ direction. The hopping matrices $\hat{t}_1$, $\hat{t}_2$, $\hat{t}_A$, and $\hat{t}_B$ are proportional to the identity in spin space, corresponding to nearest- and next-nearest-neighbor hopping amplitudes on alternating bonds and sites, respectively.

The onsite potentials $\epsilon_n$ follow a deterministic Fibonacci sequence generated recursively from $F_0=B$ and $F_1=A$, where $A$ and $B$ denote two distinct atomic species with onsite energies $\epsilon_A$ and $\epsilon_B$. To enable controlled tuning, the nearest-neighbor hopping amplitudes are parametrized as
\begin{equation}
\begin{split}
t_1 = t(1+ \delta \cos\beta), \\
t_2 = t(1-\delta \cos\beta),
\end{split}
\label{eq6}
\end{equation}
while the next-nearest-neighbor hopping amplitudes are chosen as
\begin{equation}
\begin{split}
t_A = g_a + \eta \cos (\alpha+\beta), \\
t_B = g_b + \eta \cos (\alpha-\beta).
\end{split}
\label{eq6}
\end{equation}

The source and drain electrodes are modeled as semi-infinite nonmagnetic tight-binding chains described by
\begin{equation}
H_S = \bf \sum_{n\leq -1} a^{\dagger}_{n}\epsilon_{0}a_{n}
+\sum_{n\leq -1} \left(a^{\dagger}_{n}t_0a_{n-1}+a^{\dagger}_{n-1}t_0a_{n}\right),
\label{eq6}
\end{equation}
and
\begin{equation}
H_D = \bf \sum_{n\geq 1} b^{\dagger}_{n}\epsilon_{0}b_{n}
+\sum_{n\geq 1} \left(b^{\dagger}_{n}t_0b_{n+1}+b^{\dagger}_{n+1}t_0b_{n}\right),
\label{eq7}
\end{equation}
where $t_0$ and $\epsilon_0$ denote the nearest-neighbor hopping amplitude and onsite energy of the electrodes, respectively.

The tunneling Hamiltonian couples the source (drain) electrode to the first (last) site of the SSH chain and is given by
\begin{equation}
H_{tn} = \bf \left(a_{-1}^{\dagger}t_s c_1 + c_m^{\dagger}t_db_1 + h.c.\right),
\label{eqtn}
\end{equation}
where $t_s$ and $t_d$ quantify the coupling strengths between the chain and the source and drain electrodes.
\vskip 0.1cm
\noindent
\emph{\textbf{Spin-Resolved Transmission Characteristics:}}
Spin-dependent transport is analyzed within the Green’s function formalism~\cite{negf1,negf2}. The retarded (advanced) Green’s function of the central region is
\begin{equation}
G^r=(G^a)^\dagger =[EI-H_R-\Sigma_S-\Sigma_D]^{-1},
\label{eq8}
\end{equation}
where $\Sigma_S$ and $\Sigma_D$ are the source and drain self-energies, encoding open-boundary conditions and lead-induced broadening.

The spin-resolved transmission probability is given by
\begin{equation}
\tau^{\sigma\sigma^\prime}=\text{Tr}\!\left[\Gamma_S^\sigma G^r \Gamma_D^{\sigma^\prime} G^a\right],
\label{eq9}
\end{equation}
with $\sigma=\sigma^\prime$ ($\sigma\neq\sigma^\prime$) corresponding to spin-conserving (spin-flip) processes. The coupling matrices are obtained from
\begin{equation}
\Gamma_{S(D)}^{\sigma\sigma^\prime}
=i\!\left[\Sigma_{S(D)}^{\sigma\sigma^\prime}
-\left(\Sigma_{S(D)}^{\sigma\sigma^\prime}\right)^\dagger\right].
\label{eq10}
\end{equation}
The effective transmission in each spin sector is then
\begin{equation}
T_{\uparrow}=\tau^{\uparrow\uparrow}+ \tau^{\downarrow\uparrow},
\hspace{0.5cm}
T_{\downarrow}=\tau^{\downarrow\downarrow}+ \tau^{\uparrow\downarrow}.
\label{eq11}
\end{equation}
The spin polarization is defined in terms of the spin-resolved transmission probabilities as
\begin{equation}
P_{\sigma}(E)=\frac{T_{\uparrow}(E)-T_{\downarrow}(E)}
{T_{\uparrow}(E)+T_{\downarrow}(E)},
\label{eq:polarization}
\end{equation}

\paragraph*{Results.-}
Based on the theoretical framework outlined above, we now present an analysis of the numerical results. Throughout this work, the electronic temperature is fixed at absolute zero. For simplicity, we adopt natural units by setting $c=h=e=1$. Unless stated otherwise.
\begin{figure}[ht]
{\centering \resizebox*{7.5cm}{7.5cm}{\includegraphics{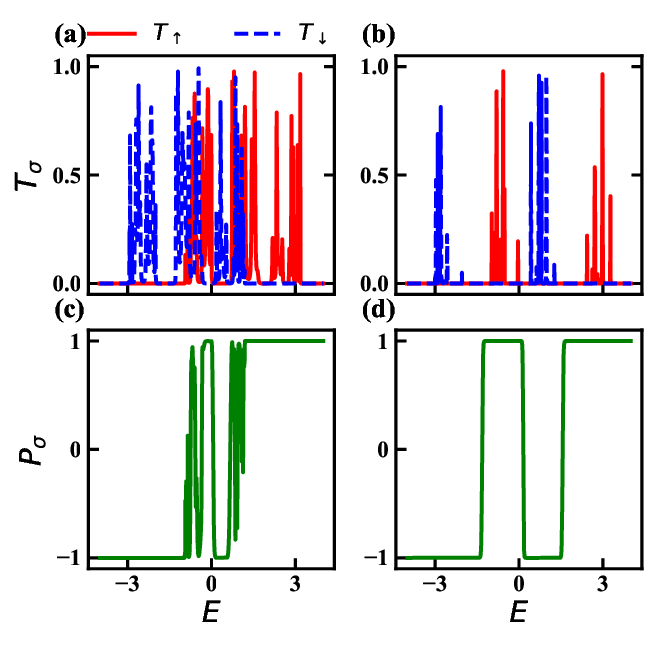}}\par}
\caption{(Color online). Energy-dependent transmission and spin polarization for two representative NN-hopping configurations.
(a) Spin-resolved transmission showing partial overlap of up- and down-spin channels.
(b) Complete separation of spin-resolved transmission within the NN-hopping framework.
(c) Corresponding spin polarization for (a), exhibiting energy-dependent fluctuations due to channel overlap.
(d) Corresponding spin polarization for (b), yielding quantized values $P=\pm1$ due to complete spin-channel separation.}
\label{fig2}
\end{figure}

We first analyze the energy-dependent transmission function $T_{\sigma}(E)$, focusing on controlled tuning of nearest-neighbor (NNH) and next-nearest-neighbor hopping (NNNH) parameters. We begin with an SSH chain containing only intercell and intracell NNH amplitudes $t_1$ and $t_2$, parametrized as $t_1=t(1+\delta\cos\beta)$ and $t_2=t(1-\delta\cos\beta)$. Within this minimal model, a overlapping between up- and down-spin transmission channels emerges, as shown in Fig.~\ref{fig2}(a), where complete spin separation is obtained for $\delta=0.8$ and $\beta=\pi/5$, while in Fig.~\ref{fig2}(b) complete separation occurs for $\delta=0.5$ and $\beta=\pi/2$, with all other parameters kept fixed. The onsite energies follow a Fibonacci sequence with values $\pm0.5$~eV. Results are shown for a Fibonacci chain of order $8$ ($L=34$), with higher-order systems discussed later. The transmission profiles demonstrate that simultaneous tuning of inter and intracell nearest-neighbor hopping, particularly via cosine modulation, provides efficient control over spin-channel separation and spin-dependent transport.
\begin{figure}[ht]
{\centering \resizebox*{8.2cm}{4.4cm}{\includegraphics{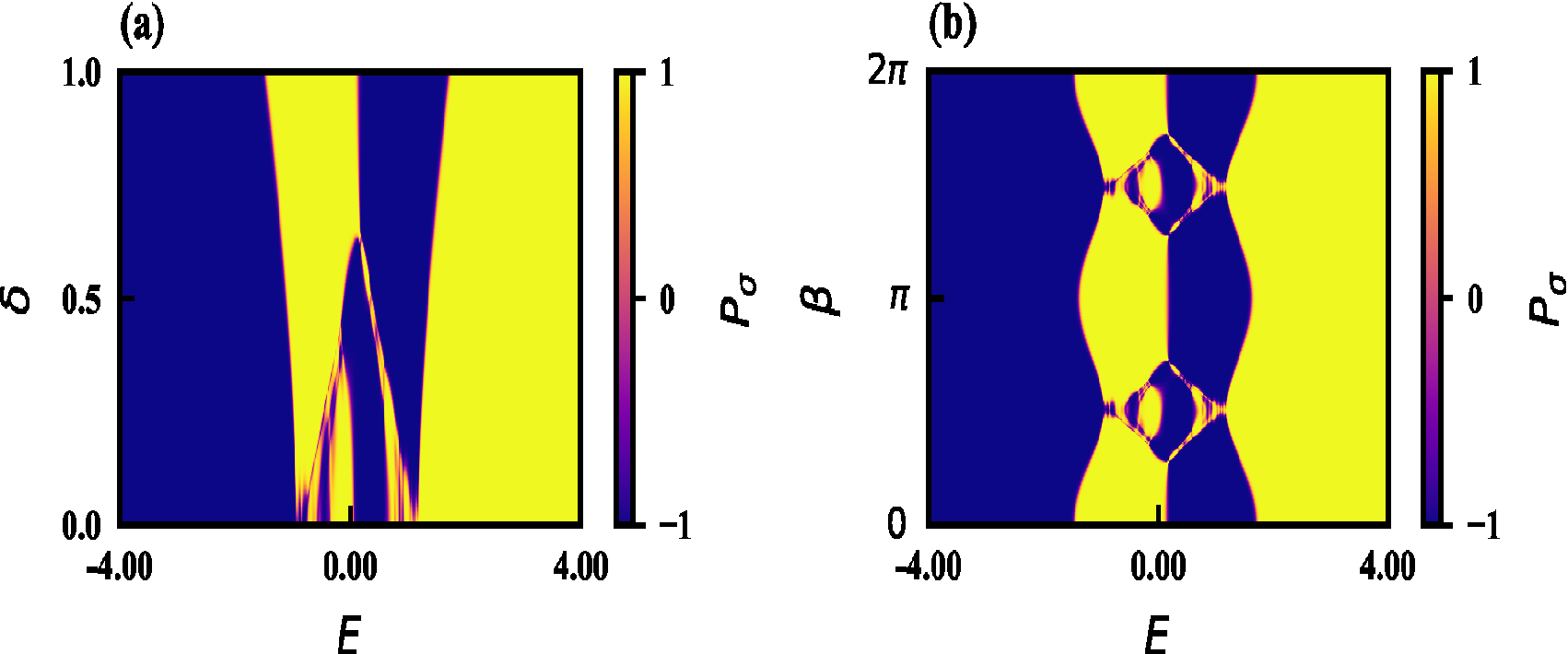}}\par}
\caption{(Color online). Phase diagrams of the spin polarization $P_\sigma$.
(a) Energy–$\delta$ dependence, showing predominantly quantized values $P_\sigma=\pm1$ over most of the parameter space, with only small regions of intermediate polarization.
(b) Energy–$\beta$ dependence, again exhibiting near-complete spin polarization over a broad parameter range, with a clear periodic modulation along the $\beta$ axis arising from the cosine-modulated hopping amplitudes. }
\label{fig3}
\end{figure}
We next evaluate the spin polarization $P_{\sigma}(E)$ using Eq.~\ref{eq:polarization}. For the parameter set of Fig.~\ref{fig2}(a), partial overlap between the up- and down-spin transmission spectra leads to pronounced energy-dependent fluctuations in $P_{\sigma}(E)$, although complete spin polarization ($P_{\sigma}=\pm1$) is attained at specific energies where transport occurs through a single spin channel as shown in Fig.~\ref{fig2}(c). In contrast, for the parameter regime of Fig.~\ref{fig2}(b), the spin-resolved transmission channels are fully separated across the entire energy range, resulting in a quantized and energy-independent polarization. The sign of $P_{\sigma}(E)$ follows the convention of Eq.~\ref{eq:polarization}, with positive (negative) values corresponding to dominant up-spin (down-spin) transmission as shown in Fig.~\ref{fig2}(d). These results establish that complete separation of spin-resolved transmission spectra is sufficient to achieve perfect spin polarization over broad energy windows.

\begin{figure}[ht]
{\centering \resizebox*{7.5cm}{7.5cm}{\includegraphics{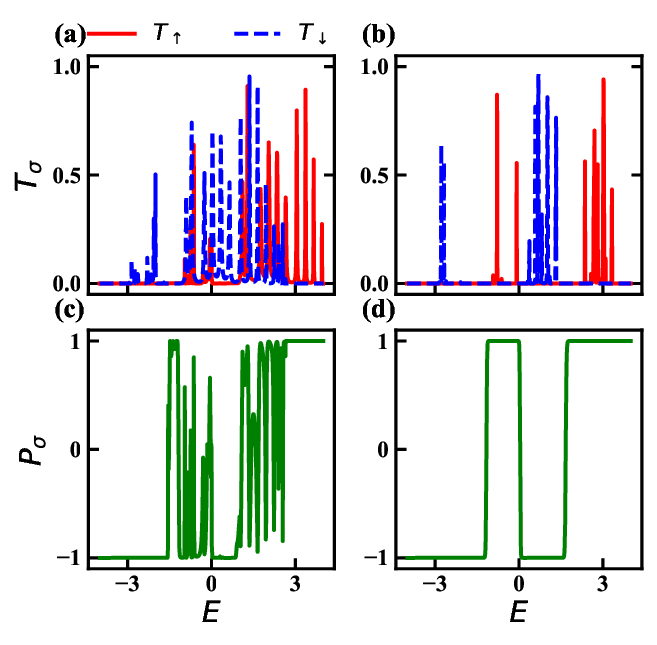}}\par}
\caption{(Color online). Energy-dependent transmission and spin polarization for two representative NNN-hopping configurations.
(a) Spin-resolved transmission with partial overlap of up- and down-spin channels.
(b) Complete separation of spin-resolved transmission within the NNN-hopping framework.
(c) Corresponding spin polarization for (a), showing energy-dependent fluctuations due to channel overlap.
(d) Corresponding spin polarization for (b), yielding quantized values $P_{\sigma}=\pm1$ as a result of complete spin-channel separation.}
\label{fig4}
\end{figure}
\begin{figure*}[ht]
{\centering \resizebox*{15cm}{10.0cm}{\includegraphics{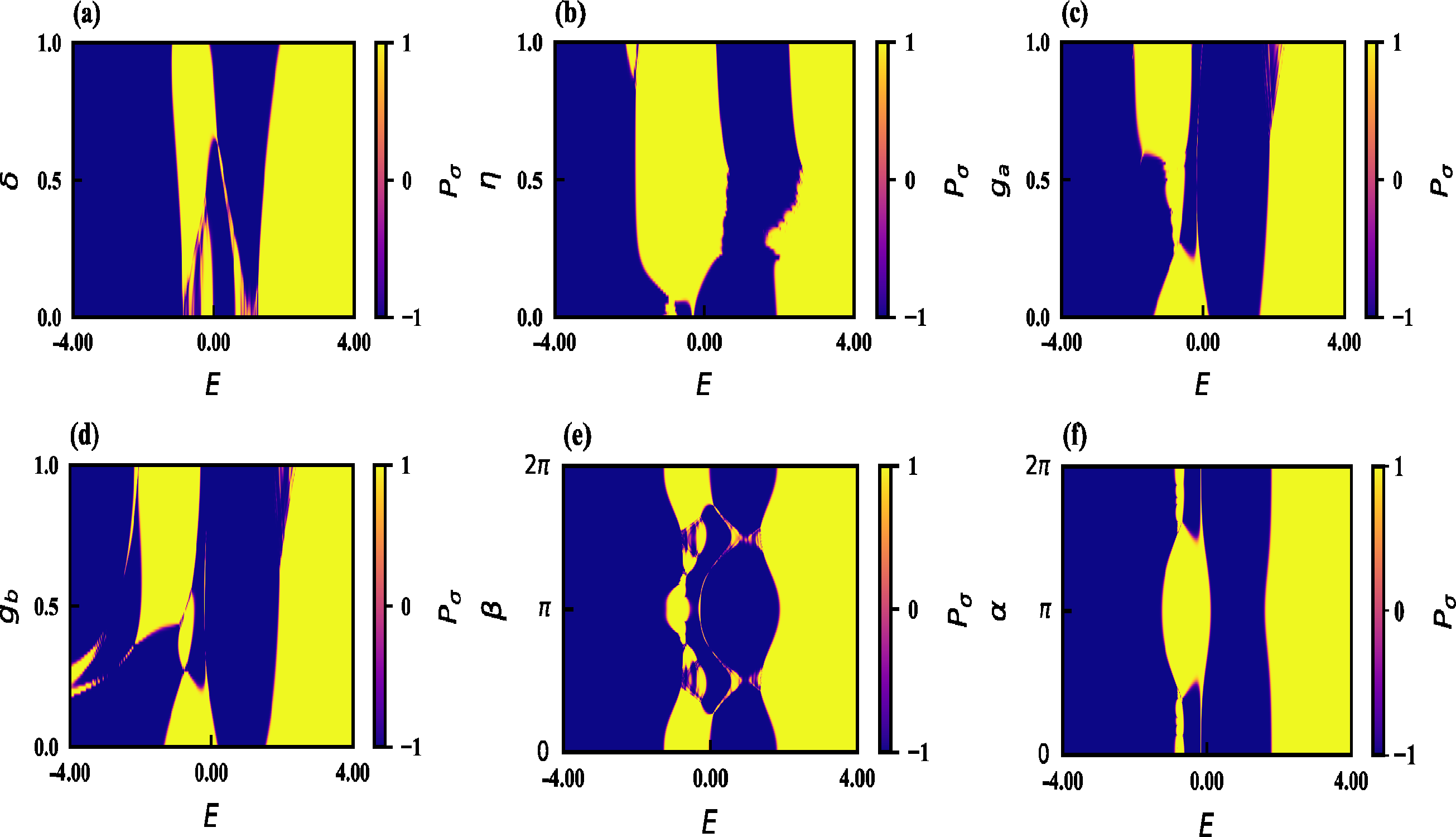}}\par}
\caption{(Color online). Phase diagrams of the spin polarization $P$ in the presence of both NN and NNN hopping.
(a),(b) Energy–$\delta$ and energy–$\eta$ dependence, showing predominantly quantized values $P=\pm1$ over most of the parameter space, with only small regions of intermediate polarization.
(c),(d) Energy–$g_a$ and energy–$g_b$ dependence, again revealing near-complete spin polarization across broad parameter ranges.
(e),(f) Energy–$\beta$ and energy–$\alpha$ dependence of the polarization a clear periodic modulation along the $\beta$ axis is observed due to cosine-modulated hopping, while no such periodicity appears in the $\alpha$ dependence.}
\label{fig5}
\end{figure*}
To elucidate the role of the hopping modulation parameters $\delta$ and $\beta$ on spin polarization, we construct phase diagrams of $P_{\sigma}(E)$ while keeping all other parameters fixed to those of Fig.~\ref{fig2}(b). As shown in Fig.~\ref{fig3}(a), the polarization predominantly attains quantized values $P_{\sigma}(E)=\pm1$ over most of the $(E,\delta)$ parameter space, with only narrow regions of intermediate polarization arising from residual overlap of spin-resolved transmission channels. The sign of $P(E)$ reflects whether transport is dominated by up- or down-spin electrons, while its magnitude remains unity whenever one spin channel is fully suppressed. This behavior demonstrates that complete spin polarization persists over broad energy windows and a wide range of $\delta$, highlighting strong robustness without the need for fine-tuning. Fig.~\ref{fig3}(b) shows the dependence of $P_{\sigma}(E)$ on the modulation parameter $\beta$, where a clear periodic pattern emerges as a direct consequence of the cosine-modulated hopping amplitudes. Similar to the $\delta$-dependent case, $P_{\sigma}(E)=\pm 1$ dominates most of the $(E,\beta)$ parameter space, with intermediate values confined to small regions due to partial channel overlap. The prevalence of fully polarized regions confirms that tuning $\beta$ provides an efficient and robust route to complete spin polarization over broad energy windows, underscoring the flexibility of the proposed mechanism for spintronic and spin-caloritronic applications.


To achieve finer control over the transmission characteristics, we next incorporate NNNH processes, parametrized as $t_M=g_a+\eta\cos(\alpha+\beta)$ and $t_N=g_b+\eta\cos(\alpha-\beta)$ for hopping between odd and even sites, respectively. In addition to enhancing tunability, NNN hopping provides a more realistic description, as effective inter-orbital hopping in real materials is often captured by longer-range terms. The introduction of four independent parameters significantly enhances the tunability of the NNNH amplitudes and, consequently, the spin-resolved transmission. For the parameter set $\delta=0.8$, $\beta=0.7\pi$, $g_a=g_b=0.5$, $\alpha=3\pi/4$, and $\eta=0.5$. of Fig.~\ref{fig4}(a), the spin-resolved transmission spectra exhibit significant overlap and  in Fig.~\ref{fig4}(b), well-separated spin channels are obtained for $\delta=0.8$, $\beta=\pi/5$, $g_a=g_b=0.1$, $\alpha=3\pi/4$, and $\eta=0.1$. thus appropriate modification of the NNN hopping parameters leads to complete separation of the spin channels, similar to the behavior observed earlier in Fig.~\ref{fig2}(b). Correspondingly, the spin polarization shows strong energy-dependent oscillations when the channels overlap [Fig.~\ref{fig4}(c)], but exhibits a square-wave–like profile with quantized values $P_{\sigma}(E)=\pm1$ over the entire energy range once the channels are fully separated [Fig.~\ref{fig4}(d)], indicating complete and robust spin polarization with the sign determined by the dominant spin channel. These results establish NNN hopping as an effective and independent control parameter for spin-dependent transport in the extended SSH model, providing additional tunability that complements nearest-neighbor modulation and offering a promising route for realizing controllable spintronic functionalities.

We next examine the spin polarization as a function of energy and parameters associated with next-nearest-neighbor (NNN) hopping by constructing phase diagrams of $P_{\sigma}(E)$, varying one parameter at a time while keeping the others fixed to the values of Fig.~\ref{fig4}(b). Figure~\ref{fig5}(a) shows the dependence on the modulation amplitude $\delta$, where the phase diagram closely resembles that of Fig.~\ref{fig3}(a), demonstrating that $\delta$ remains an efficient control parameter for spin polarization even in the presence of NNN hopping. Over most of the $(E,\delta)$ parameter space, the polarization is quantized at $P_{\sigma}(E)=\pm1$, with only narrow regions of intermediate values arising from residual overlap of spin-resolved transmission channels. Figure~\ref{fig5}(b) illustrates the effect of the NNN hopping modulation strength $\eta$, revealing that $P_{\sigma}(E)$ remains predominantly quantized over a broad range of energies and $\eta$, establishing $\eta$ itself as an effective tuning knob for robust spin polarization.

We further analyze the role of the NNN hopping offset parameters $g_a$ and $g_b$, shown in Figs.~\ref{fig5}(c) and ~\ref{fig5}(d). Although both enter the NNN hopping amplitudes in an analogous form, they produce qualitatively different effects on spin polarization, with variations in $g_b$ influencing transport more strongly than comparable changes in $g_a$. This asymmetry indicates that intercell and intracell NNN hopping processes affect spin-dependent transport in distinct ways, highlighting the nontrivial role of longer-range hopping. Nevertheless, both parameters yield robust spin polarization, with $P_{\sigma}(E)=\pm1$ dominating wide energy ranges and only small isolated regions showing partial polarization due to channel overlap.

Finally, we examine the influence of the phase angles $\beta$ and $\alpha$ [Figs.~\ref{fig5}(e) and ~\ref{fig5}(f)]. Consistent with Figs.~\ref{fig3}(b), the $(E,\beta)$ phase diagram exhibits a clear periodic structure originating from the cosine modulation of both nearest- and next-nearest-neighbor hopping amplitudes, with $\beta$ acting as a modulation phase. As a result, $P_{\sigma}(E)$ remains predominantly quantized over most of the $(E,\beta)$ parameter space. In contrast, the $(E,\alpha)$ phase diagram shows no apparent periodicity, reflecting the distinct manner in which $\alpha$ enters the effective hopping parameters; nevertheless, $P_{\sigma}(E)=\pm1$ persists over large regions, with only isolated areas of partial polarization.

Taken together, these results underscore the highly nontrivial and complementary roles of nearest- and next-nearest-neighbor hopping processes in shaping spin-dependent transport in the extended SSH model. The independent tunability of intercell and intracell NN and NNN hoppings provides enhanced flexibility for spin manipulation, enabling robust spin polarization and establishing the system as a versatile platform for exploring spin-selective transport phenomena in engineered low-dimensional structures.

\paragraph*{Conclusion.-}
In summary, we have studied spin-dependent quantum transport in an extended Su--Schrieffer--Heeger model featuring cosine-modulated nearest- and next-nearest-neighbor hopping, employing the nonequilibrium Green's function formalism at zero temperature. We find that an appropriate tuning of the nearest-neighbor hopping parameters induces a complete separation of spin channels, leading to perfect spin polarization over wide energy windows, while the inclusion of next-nearest-neighbor hopping enhances tunability and robustness. Systematic phase-diagram analyses reveal that quantized spin polarization (\(P_{\sigma}=\pm1\)) emerges across extended regions of parameter space rather than at isolated fine-tuned points, with intercell and intracell hopping processes exerting distinct and asymmetric influences on spin transport. The persistence of these features with increasing system size further underscores their robustness, thereby establishing the extended SSH model as a versatile platform for controllable spin-polarized transport with potential implications for spintronic and spin-caloritronic applications.

\section*{End Matter}
\section{Finite-Size Robustness}
\begin{figure}[ht]
{\centering \resizebox*{7.5cm}{7.5cm}{\includegraphics{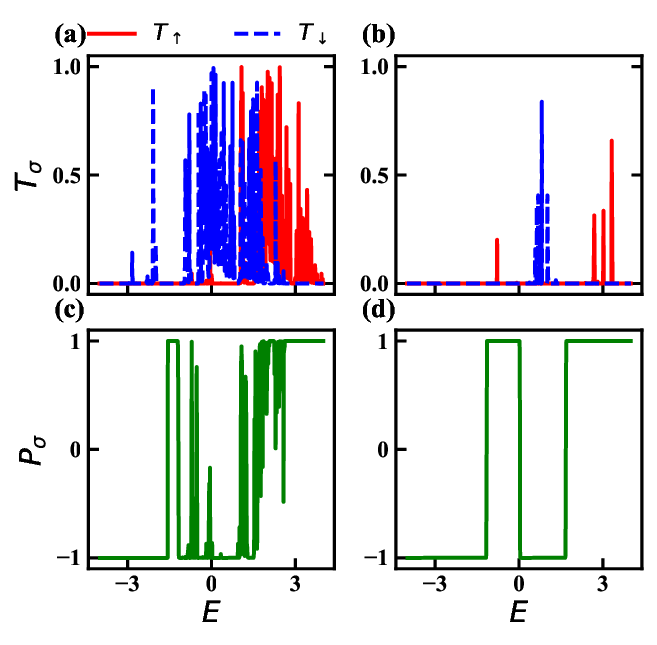}}\par}
\caption{(Color online). Same as Fig.~4, but for a higher-order Fibonacci chain of generation 11 ($L=144$).}
\label{fig6}
\end{figure}
We further examine the robustness of our results by considering a higher-order system, namely a Fibonacci chain of order $11$ consisting of $144$ sites. Employing the same nearest- and next-nearest-neighbor hopping parameters as used in Fig.~4, we compute the spin-resolved transmission and observe that, even for this substantially larger system size, a suitable choice of hopping parameters results in a complete separation of the up- and down-spin channels, yielding perfect spin polarization. The persistence of this behavior with increasing system size confirms that the observed spin polarization is not a finite-size artifact but an intrinsic characteristic of spin-dependent transport in the extended SSH model with modulated hopping. This highlights the robustness of the proposed mechanism and underscores its potential relevance for realistic mesoscopic and nanoscale implementations.

\end{document}